\documentclass[11pt]{article}
\usepackage{epsfig}

\usepackage{amssymb,amsfonts}

\newtheorem{theorem}{Theorem}

\newenvironment{proof}{{\noindent \it Proof:\/}}{$\square$}

\newcommand{\braced}[1]{{\left\{ #1  \right\} }}
\newcommand{\ie}{{\em i.e.\/}}
\newcommand{\cost}{{\mbox{\it cost\/}}}
\newcommand{\opt}{{\mbox{\scriptsize \sc opt}}}
\newcommand{\optcost}{{\mbox{{\it optcost\/}}}}

\newcommand{\leftbracedthree}[3]{\left\{\begin{array}{l}
                                        #1 \\ #2 \\ #3
                                    \end{array}\right.
                           }

\newcommand{\calA}{{\cal A}}
\newcommand{\calB}{{\cal B}}

\newcommand{\calD}{{\cal D}}

\newcommand{\calJ}{{\cal J}}

\begin{document}

\title{A Program to Determine the Exact Competitive Ratio
of  List $s$-Batching with Unit Jobs
\footnote{The research of the first and third authors was
supported by NSF grant CCR-0132093.}}

\author{
Wolfgang Bein \thanks{School of Computer Science, 
                         University of Nevada, 
			 Las Vegas, NV 89154, USA. 
			 \tt{bein@cs.unlv.edu}} \and 
Leah Epstein \thanks{Department of Mathematics, 
			University of Haifa, 
			31905 Haifa, Israel.
			{\tt lea@math.haifa.ac.il}} \and 
Lawrence L. Larmore \thanks{{School of Computer Science, 
			University of Nevada,
			Las Vegas, NV  89154, USA. 
			\tt{larmore@cs.unlv.edu}}}
\and John Noga \thanks{Department of Computer Science, 
			California State University,
			Northridge, CA 91330, USA. \tt{jnoga@csun.edu}}}

\date{}

\maketitle

\begin{abstract}
We consider the online list s-batch problem, where all the jobs have processing
time 1 and we seek to minimize the sum of the completion 
times of the jobs. We give a Java program which is used to verify
that the competitiveness of this problem is $\frac{619}{583}$.

\medskip

\noindent
{\em Keywords:} Design of Algorithms; Online Algorithms;
Batching; TCP acknowledgment.

\end{abstract}


\section{Background}

In the paper "Optimally Competitive List Batching" \cite{TCS09} we give
results regarding online batching problems. For one
problem -- the  online list s-batch problem, where all the jobs 
have processing time 1 and we seek to minimize the sum of the completion
times of the jobs -- we have used a computer program to obtain some
of our results. The purpose of this document to make this
program publicly available. The program is printed 
in Section \ref{sec: program}. It is also available for
download from www.egr.unlv.edu/$\sim$bein/pubs/VerifyLowerBound.java in ASCII format. 
The reader should consult
the full paper \cite{TCS09} or the earlier conference version
\cite{SWAT04}, but in the interest of self-containedness,  
we briefly define the problem
in Section \ref{sec: introduction} and repeat the results
in Section \ref{sec: results}.

\section{Introduction}\label{sec: introduction}
A {\em batching problem\/} is a scheduling problem where a set of
jobs $\calJ = \{ J_i \}$ with processing times $\braced{p_i\ge0}$
must be scheduled on a single machine, and where
$\calJ$ must be partitioned into {\em batches\/}
$\calB_1, \ldots,  \calB_r$.
All jobs in the same batch are run jointly and each job's completion time
is defined to be the completion time
of its batch.  We assume that when a batch is scheduled it requires a
setup time $s$.
In an {\em s-batch} [sequential] problem the {\em length\/}
of a batch, \ie, the time required to process the batch,
is the sum of the processing times of its member jobs.
The goal is to find a schedule that minimizes the
{\em sum of completion times\/} $\sum C_i$, where $C_i$
denotes the completion time of $J_i$ in a given schedule.

Given a sequence of jobs, a batching algorithm must
assign every job $J_i$ to a batch. More formally, a feasible
solution is an assignment of each job $J_i$ to the $m_i^{th}$
batch, $i \in \{1,\ldots, n \}$.
In this paper, we consider the {\em list\/} version of the problem,
where the given order of the jobs must be respected,
\ie, $m_i\le m_j$ if $i < j$.

An {\em online\/} algorithm for a list batching problem
must choose each $m_i$ before receiving $J_{i+1}$,
\ie, each job must be scheduled before a new job is seen,
and even before knowing whether $J_i$ is the last job.
After receiving $J_i$, an algorithm has only two choices, namely
whether to assign $J_i$ to the same batch as $J_{i-1}$ or not.
Throughout this paper, we will use the phrase ``$\calA$ batches
at step $i$" to mean that algorithm $\calA$ decides
that $J_i$ is the first job of a new batch, {\ie}
$m_i = m_{i-1}+1$.  We use the phrase ``current batch" to denote
the batch to which the last job was assigned.
Then, when $J_i$ is received, $\calA$ must decide whether to
add $J_i$ to the current batch, or ``close"
the current batch and assign $J_i$ to a new batch.

Online algorithms are analyzed in terms of {\em competitiveness\/},
a measure of the performance that compares the decision made online
with the optimal offline solution for the same problem.
We say that an algorithm $\calA$ is {\em $C$-competitive\/}
if, for any sequence of jobs $\braced{J_i}$, $\calA$ finds a schedule
whose cost is at most $C\cdot\cost_\opt$, where
$\cost_\opt$ is the minimum cost of any schedule for
that input sequence.

\section{The Case $s=1$ and $p_i =1$}\label{sec: results}

In our papers \cite{TCS09,SWAT04} we give a solution for the 
offline problem.
We define a function $F[n]$ for $n \geq 0$, as follows.
For $n = m(m+1)/2 + k$ for some $m \geq 0$ and some $0 \leq k \leq m+1$, then
\begin{equation}
F[n] = \frac{m(m+1)(m+2)(3m+5)}{24} + k(n+m-k+1) + \frac{k(k+1)}{2}.
\end{equation}

\begin{theorem}\label{thm: optcost is F}
For $\optcost[n]$, $\optcost = F[n]$ for all $n \geq 0$.
Furthermore, if
$n = \frac{m(m+1)}{2} + k$ for some $m \geq 0$ and some $0 \leq k \leq m+1$,
the optimal size of the first batch is\\
$\leftbracedthree{m \mbox{\rm\ if\ } k = 0}
{m \mbox{\rm\ or\ } m+1 {\rm\ if\ }0 < k < m+1}
{m+1 \mbox{\rm\ if\ } k = m+1}$\\
\end{theorem}

For the online problem we define the following algorithm in \cite{SWAT04,TCS09}.
Define $\calD$ to be the online algorithm which batches after jobs:
2, 5, 9, 13, 18, 23, 29, 35, 41, 48, 54, 61, 68, 76, 84, 91,
100, 108, 117, 126, 135, 145, 156, 167, 179, 192, 206, 221,
238, 257, 278, 302, 329, 361, 397, 439, 488, 545, 612, 690,
781, 888, 1013, 1159, 1329, 1528, 1760, and
2000+40$i$ for all $i \ge 0$.

\begin{theorem}\label{thm: identical 619 583}
$\calD$ is $\frac{619}{583}$-competitive,
and no online algorithm for
the list s-batch problem restricted to unit job sizes
has competitiveness smaller than $\frac{619}{583}$.
\end{theorem}
\begin{proof}
Consider the algorithm $\calD$ described above.
Verifying that $\calD$ maintains a cost ratio of at most $\frac{619}{583}$
for all job sequences with less than 2000 jobs is done by the Java program
in Section \ref{sec: program}.
Consider now sequences with more than 2000 jobs.
Then the contribution of job $i > 2000$ to the optimal cost is
at least $i+1$. For $\calD$ the contribution of this job $i$ consists of
the setup times prior to job $2000$ plus the
setup times later, plus the number jobs  $i$ as well the the subsequent
jobs in the same batch. This amount is no more than
$48 + \lceil\frac{i-2000}{40}\rceil+ i + 39$, because 39 is the
maximum number of additional jobs in this batch.
Thus the for the ratio of cost we obtain
\begin{eqnarray*}
\frac{48 + \lceil\frac{i-2000}{40}\rceil+ i + 39}{i+1} &\le&
\frac{\frac{41}{40} i + 38}{i+1}
\;<\;
\frac{41}{40} + \frac{38}{i+1}  \;\le\;
\frac{619}{583}
\end{eqnarray*}

The contribution of the first 2000 jobs to the optimal cost
is larger than a contribution in a short sequence (with 2000 jobs or less)
because the size of the optimal batches increase with the
number of jobs.
Therefore $\calD$ is $\frac{619}{583}$-competitive.

We now turn to the verification of the lower bound. Any online algorithm
for list batching restricted to unit jobs
is described by a sequence of decisions: should the $i^{th}$
job be the first job in a new batch? Thus any such online algorithm
can be represented as a path in a decision tree where a node at
level $i$ has two children: one representing the choice not
to batch prior to job $i$ and one representing making job $i$
the first job in a new batch.  We note that the algorithm never
batches upon the arrival of the first job.
We have verified that any path from the root to a node with depth
$d$ in this decision tree must encounter a node at which the
ratio of online cost to offline cost is at least $\frac{619}{583}$;
and thus we have established that lower bound.  Verification was
done using our computer program.

\end{proof}

Interestingly, the lower bound verification program requires consideration
of only the portion of the decision tree to depth 100.  That is, if the
decision tree is truncated at any level less than 100, the lower bound is not
obtained.  What this means is that, if an online algorithm is informed in
advance that there will be at most 99 jobs, it can achieve a competitiveness
less than $\frac{619}{583}$.

\begin{figure}[ht]
\begin{center}
\includegraphics[width = 5.0in]{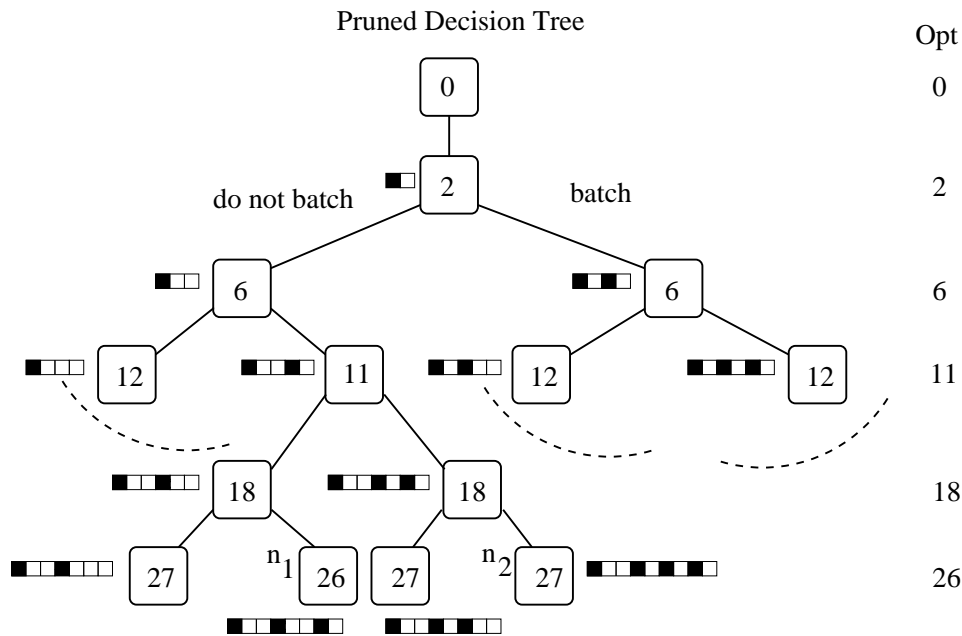}
\end{center}
\caption{The Decision Tree used in the Pruning Procedure}
\label{fig: decision}
\end{figure}

Given that there are exponentially many paths from the root
to a node at depth $d$, two notes on efficiency are appropriate
here. First, if a node is encountered where the ratio of costs is
greater than or equal to $\frac{619}{583}$ then no further descendants
need to be checked. This alone brings the calculation described
above to manageable levels. Second, given two nodes $n_1$ and $n_2$
which have not been pruned by the previous procedure, if the online
cost at $n_1$ is less or equal to the online cost at $n_2$ and
both have done their most recent batching at the same point then
descendants of $n_2$ need not be considered. This follows because
the cost on any sequence of choices leading from $n_2$ is greater
or equal to the same cost on $n_1$.
We illustrate the preceding ideas with the diagram of Figure
\ref{fig: decision}.
Level $i$ contains all possible decisions after $i$ jobs have
arrived. The symbolic Gantt chart next to every decision node
show the schedule the algorithm constructs at that node. In the Gantt charts
black squares denote setup times, while white squares are used to denote jobs.
The cost of the algorithm is written into the node.
Note that we can prune at level 3 because
$\frac{12}{11} > \frac{619}{583}$. Also note that
descendants of node $n_2$ need not be considered.

\section{Java Program}\label{sec: program}
The program is compiled with {\tt javac}, version {\tt 1.6.0\_10}.
\begin{verbatim}

import java.util.LinkedList;
import java.util.ListIterator;
import java.util.Vector;


public class VerifyLowerBound {
   // Maximum number of jobs that should be considered
   private static int howFar=100;

   /* The idea is to keep the leaves in the tree (the partial candidates) which 
      must be considered (have not been pruned yet) in a linked list called candidates.
    */

   public static void main(String[] args) {
      int count = 0; 
      int[] opt = calculateOpt(); 
      int[] suggested = {0,2,5,9,13,18,23,29,35,41,48,54,61,68,76,84,91,100,108,117,
                         126,135,145,156,167,179,192,206,221,238,257,278,302,329,361,
                         397,439,488,545,612,690,781,888,1013,1159,1329,1528,1760,2000};
      LinkedList<PartialCandidate> candidates = new LinkedList<PartialCandidate>();
      PartialCandidate pC;

      // Start with the tree that results after considering just 1 job
      candidates.add(new PartialCandidate());

      /* While there are candidates remaining, remove the first. If it has reached the
         maximum number of jobs that should be considered, print out the solution and 
         add one to the count. If the candidate  has not reached the maximum number of 
         jobs, check whether 
         1) has a cost ratio strictly smaller than 619/583    AND
         2) it does not have a just as good candidate already in the list.
         If both are true, then add both of its children to the list of candidates.
       */

      while (candidates.size() !=0) {
         pC = candidates.remove();
         if (pC.getHowFar() > howFar) {
            count++;
            System.out.println(pC);
         }
         else if (583*pC.evalCost() < 619*opt[pC.getHowFar()]) {
            if(!existJustAsGood(pC,candidates)) {
               candidates.add(new PartialCandidate(pC,false));
               candidates.add(new PartialCandidate(pC,true));
            }
         }
      }
      System.out.print(count + " candidates achieve a cost ratio strictly smaller");
      System.out.println("than 619/583 on all sequences no longer than " + howFar);

      // Verify that the suggested candidate succeeds up to 2000 jobs
      boolean success = true;
      int value = 0, j=0;
      for (int i=1; i<2000; i++) {
         value += (i-suggested[j]-1)+i+j+1;
         if (i==suggested[j+1])
            j++;
         if (583*value > 638*opt[i])
            success = false;
      }
      System.out.println("The suggested candidate is a success up to 2000 jobs? " + success);
   }

   // Uses Theorem 1 to calculate the offline optimum costs
   private static int[] calculateOpt() {
      int n, m=0, howMany=2001;
      int[] answer = new int[howMany];
      answer[0] = 0;
      for (n=0;n<howMany-1;n++) {
         if (2*n<(m+2)*(m+1))
            answer[n+1]=answer[n]+n+m+2;
         else {
            m++;
            answer[n+1]=answer[n]+n+m+2;
         }
      }
      return answer;
   }



   /* Checks to see if the list already has a candidate which has gotten as far,
      has no greater cost, uses no more set ups, and has the most recent set up 
      at the same location, 
    */
   private static boolean existJustAsGood(PartialCandidate pC, 
                              LinkedList<PartialCandidate> ll) {
      PartialCandidate jAGC;
      ListIterator<PartialCandidate> i = ll.listIterator(0);
      while (i.hasNext()) {
         jAGC = i.next();
         if (
             jAGC.getHowFar() == pC.getHowFar() &&
             jAGC.evalCost() <= pC.evalCost() &&
             jAGC.getSetUps().size() <= pC.getSetUps().size() &&
             jAGC.getSetUps().lastElement().equals(pC.getSetUps().lastElement())
            )
            return true;
      }
      return false;
   }
}


/* An algorithm for the unit job batching problem can be specified by
   where the set up times occur (or equivalently where it batches). A 
   partial candidate is an incompletely specified algorithm. A partial 
   candidate has a list (stored in setUps) which specifies where the 
   set up times occur only for sequences of up to a certain length (ie
   the depth in the tree - stored in howFar).  
 */


class PartialCandidate {

   private Vector<Integer> setUps;
   private int howFar;

   // Create the only reasonable candidate for the sequence of 1 job.
   public PartialCandidate() {
      setUps = new Vector<Integer>();
      setUps.add(0);
      howFar = 1;
   }

   /* Given a partial candidate create a new partial candidate which
      is defined for sequences one job longer. Determine whether to
      add an additional set up at this point based upon the value of
      batch.
    */
   public PartialCandidate(PartialCandidate pC, boolean batch) {
      Vector<Integer> prevSetUps = pC.getSetUps();
      int prevHowFar = pC.getHowFar();

      setUps = new Vector<Integer>();
      for (int i=0; i<prevSetUps.size(); i++) {
         setUps.add(prevSetUps.get(i));
      }
      if (batch)
         setUps.add(prevHowFar);

      howFar = prevHowFar+1;
   }

   // Calculate cost of the partial candidate on a sequence of howFar jobs.
   public int evalCost() {
      int value=0;
      for (int i=1; i<setUps.size(); i++) 
         value += (i+setUps.get(i))*(setUps.get(i)-setUps.get(i-1));
      value += (setUps.size()+howFar)*(howFar-setUps.get(setUps.size()-1));
      return value;
   }

   // Used to print out the partial candidate
   public String toString() {
      String str="";
      for (int i=0; i<setUps.size(); i++)
         str += setUps.get(i)+" ";
      str += "Cost "+evalCost();
      return str;
   }

   // Accessors
   public Vector<Integer> getSetUps() {
      return setUps;
   }

   public int getHowFar() {
      return howFar;
   }

}

\end{verbatim}

\end{document}